\documentclass[aps,twocolumn,showpacs,prl]{revtex4}
\usepackage{epsfig}
\usepackage{amssymb,amsfonts,amsmath,wrapfig,mathrsfs,mathbbol}
\usepackage{color}

\begin{document}

\title{Spatial entanglement in optical parametric oscillators with photonic
crystals}

\author{Maria M. de Castro,$^{1}$ Miguel Angel Garcia March,$^{2}$ Damia Gomila,$^{1}$ and 
Roberta Zambrini$^{1}$}

\affiliation{$^{1}$Institute for Cross-Disciplinary Physics and Complex Systems, IFISC
(CSIC-UIB), Palma de Mallorca, 07122, Spain}

\affiliation{$^{2}$ Department of Physics, Colorado School of Mines, Golden, CO, 80401 }
\date{\today}

\begin{abstract}
The effects of an intracavity photonic crystal in a multimode optical
parametric oscillator are  studied,  with a special focus on quantum fluctuations. 
The capability to either stimulate or inhibit the spatial instability, lowering or increasing the
parametric threshold, allows to control  the intensity fluctuations and correlations.
A significative quantum noise reduction and an
increase of the range of squeezed quadratures are found above threshold where spatial
Einstein-Podolsky-Rosen entanglement and inseparability are found.
\end{abstract}

\maketitle


Photonic crystals (PC) are dielectric media with periodic modulation of the refractive index
which can lead to gaps in the allowed frequencies of electromagnetic waves~\cite{Joannopoulos}.
Seminal works predicted the possibility to control spontaneous emission by PC~\cite{PC87} when
suppressing a  radiative transition with frequency  within the  photonic band-gap. After two
decades of intense experimental activity, these engineered media provide an unprecedented control
of light confinement, guiding, and propagation 
~\cite{Joannopoulos,BuschRussell}.  Recently, inhibition of
spontaneous emission~\cite{Noda2007}  has been experimentally shown, improving the extraction
efficiency  of light emitting devices~\cite{Lodahl} and redistributing where needed the
corresponding energy~\cite{Noda2005}. The use of PC for environment (dissipation) engineering
is also the basis of intense research activity about non-Markovian evolution of quantum 
states~\cite{structured_reserv}.
An unexplored issue we raise here is if a PC can be used to control or even suppress  light
fluctuations
to improve quantum correlations   in the continuous variable regime~\cite{contin}
in driven devices   emitting entangled 
beams. Quantum correlated bright beams can be 
generated by
optical parametric oscillators (OPOs). These are common devices giving squeezing~\cite{squeezingOPO} and
entanglement~\cite{epr_OPO}  between modes with different polarization or frequency. In 1992
similar quantum effects were also predicted between different {\it spatial} modes~\cite{lugiato},
extending the study of quantum effects to multimode
OPO~\cite{opos,zambriniEPJD},  Kerr media~\cite{lugiato,zambrini2000}, and
second harmonic generation~\cite{shg}.  Different applications based on spatially multimode
operation have been recently demonstrated in optical switching~\cite{patterns}, quantum
imaging~\cite{qimaging}, metrology~\cite{laserpointer}, and quantum information~\cite{qinfo}. 
Considering both theoretical activity and experimental achievements, the subject of spatial
entanglement is becoming a mature research topic. 

In this Letter we explore how to control and improve multimode squeezing and entanglement by means
of an intracavity photonic crystal (PC) in a nonlinear device. The prototype system considered is a
multimode degenerate OPO, well studied both below and above threshold~~\cite{opos,zambriniEPJD},
when a PC is introduced.  Within the proposed photonic-crystal optical parametric oscillator
(PCOPO), the spatial modulation is not changing the environment spectrum of fluctuations
\cite{structured_reserv} but is instead modifying the Hamiltonian of the intracavity process.  The
PC modulation takes place in the transverse plane --the device is longitudinally monomode-- and is
modeled  by a spatial profile of the otherwise homogeneous refractive index. Intracavity PC have
been predicted to allow for inhibition of the phenomenon of  pattern formation in Kerr and singly
resonant OPOs~\cite{gomilaPRL,damiaPRE} by an increase of the pump energy needed to cross the
instability threshold. The phenomenon arises when unstable spatial modes are in the band-gap
because their emission is prevented, as recently observed in two independent experimental
set-ups~\cite{expPCinib}. Here we will show that in PCOPO the possibility to tune the spatial
instability -- matching-up to the parametric threshold-- is actually wider. Indeed, the PC allows
to either increase or reduce the threshold energy due to the mixing of different frequency waves. 
Related to the instability shift is the possibility to control
 the fluorescence intensity below threshold and the spatial distribution of  quantum fluctuations.
After characterizing quantum correlations in different regimes, two major effects are identified
above threshold. First of all, the PC provides a locking mechanism to freeze the known diffusive
drift motion of the pattern~\cite{zambrini2000}  reducing by orders of magnitude the fluctuations
in the associated quadrature. A larger range of quadratures is then found to be squeezed,  decreasing
the sensitivity to the choice of the local oscillator phase in squeezing measurements. Moreover, we
find that spatial modes can be entangled above threshold when introducing the PC in the OPO, either
considering state inseparability~\cite{simon-duan} or the Einstein-Podolsky-Rosen criterion of
Ref.~\cite{reid}.


The master equation for multimode type I degenerate OPO was described in detail  by Gatti et al. 
\cite{opos}. A description valid both below and above threshold (for pump $|\alpha_0|<2$) can be obtained
through  a mapping into the Q-representation as discussed in Ref.~\cite{zambriniEPJD}. This leads to nonlinear
Langevin equations for the spatially dependent pump $\alpha_0$ and signal $\alpha_1$ fields
\begin{eqnarray} 
\partial_t
\alpha_0( x ,t)&=& - \left[(1+i\Delta_0( x ))-i\nabla^2 \right]\alpha_0(
x,t)+ \nonumber  \\ \nonumber && E-\frac{1}{2}\alpha_1^2( x ,t)+\xi_0( x ,t)
  \\\nonumber
\partial_t
\alpha_1( x ,t)&=& - \left[(1+i\Delta_1( x ))-2i\nabla^2
\right]\alpha_1( x ,t)+\\ && \alpha_0( x ,t)
\alpha_1^*( x ,t)+\xi_1( x ,t),\label{Eq:lang} 
\end{eqnarray} 
with $E$ input field, $\nabla^2$ diffraction, and  $\xi_0$ additive and $\xi_1$ multiplicative 
{\it phase-sensitive} white noises~\cite{details}.  
The PC refractive index modulation is modeled by  introducing
spatial dependent detunings $\Delta_0(x)$ and $\Delta_1(x)$, which can have different amplitudes and, in the
simplest case, have the same periodicity with wave-number $k_{pc}$. The main mechanism we aim to explore
is the effect of the band-gap on the spatially multimode down-conversion process.  As known, modulation
instability in  OPO with negative signal detuning appears at wave-number $k_c=\sqrt{-\Delta_1/2}$
\cite{oppo}. Therefore, the most  interesting configuration is for a (sinusoidal) modulation with
$k_{pc}=2 k_c$, since in this case the signal pattern would be in the photonic band-gap~\cite{gomilaPRL}.

\begin{figure}[h]
\includegraphics[width=8cm]{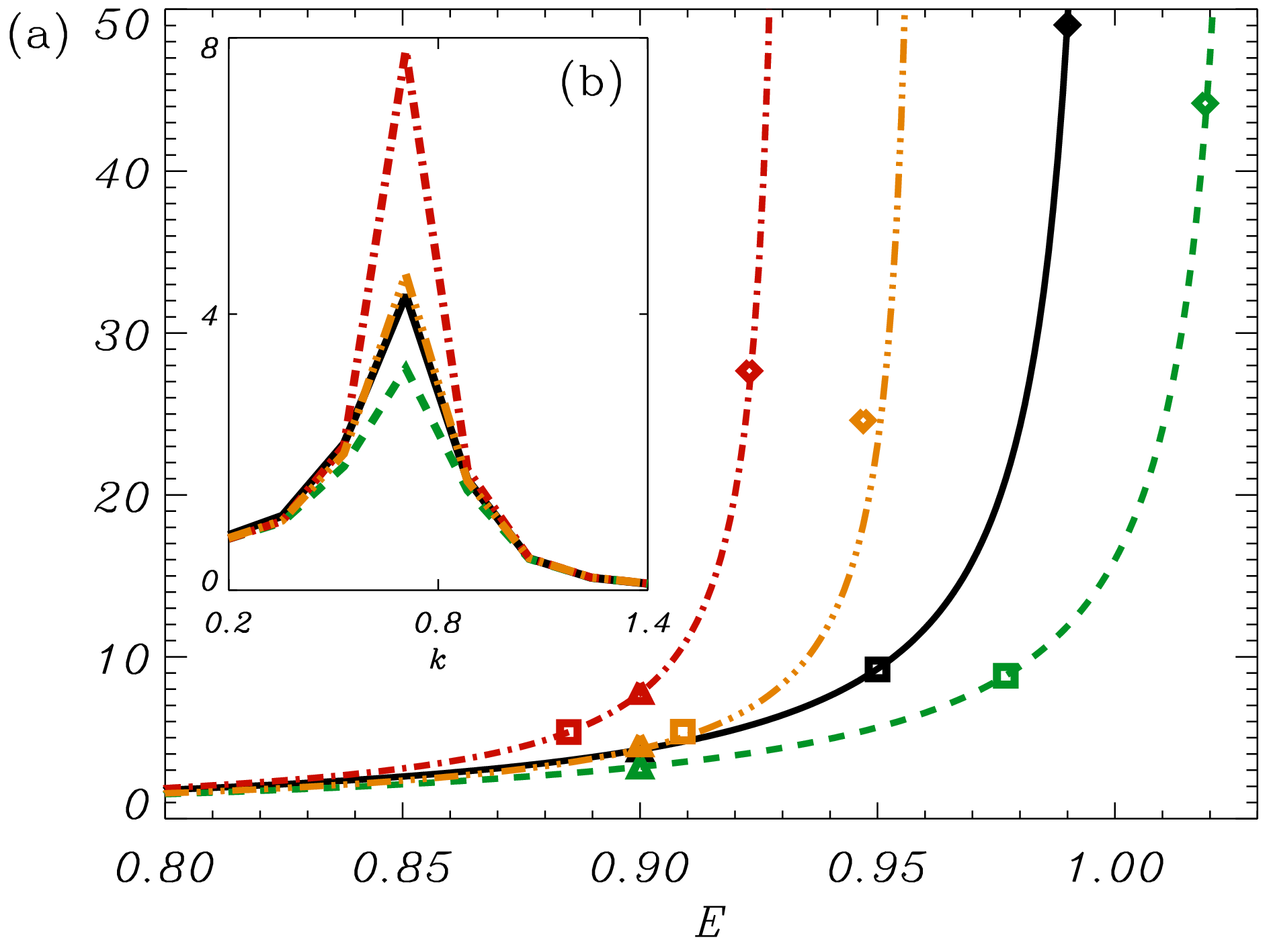} \\
\caption{(Color online)(a) Steady intensity  $\langle\hat A_1^\dagger(k_c)\hat
A_1(k_c)\rangle$ increasing with $E$ (below threshold).  Here the symbols are results of
numerical simulations of Eqs. (\ref{Eq:lang}) at different pump values, while lines
represent analytical results within a linear approximation
\cite{PCOPObelow}. (b) $\langle\hat A_1^\dagger(k)\hat A_1(k)\rangle$ from numerical 
simulations for $E=0.9$. 
Different curves correspond to the OPO without PC with 
$\Delta_0=0$ and $\Delta_1=-1$  (black solid line);
$\Delta_1=-1+0.5\sin(k_{pc}x)$ and $\Delta_0=0$ (green dashed line);
$\Delta_0=0.5\sin(k_{pc}x)$ and $\Delta_1=-1$ (red dot-dashed line);
$\Delta_0=0.5\sin(k_{pc}x)$ and $\Delta_1=-1+0.5\sin(k_{pc}x)$ (orange 3 dots-dashed
line).  \label{fig1}}
\end{figure}

We start considering   the effects on the quantum fluctuations in the PCOPO below threshold. 
On average the signal
field  vanishes everywhere $\langle\hat A_1(x)\rangle=0$ but, due to the nonlinearity
of the medium, it is not in a coherent vacuum state, either with or without the PC. When spatial
modes are considered in the far field,  the amplitude of intensity fluctuations is maximal at the
critical wavenumber~\cite{opos} (Fig.~\ref{fig1}b). As  the least
damped spatial mode falls within the PC band-gap, inhibition  of the off-axis emission is expected
\cite{PC87,Noda2007}. This effect  can be appreciated comparing the larger intensity fluctuations
for the OPO (Fig.~\ref{fig1}b, continuous line)  with respect to the PCOPO with modulation of the
signal detuning $\Delta_1(x)$ (Fig.~\ref{fig1}b, dashed line). 
On the other hand, it is rather surprising to find that, for a fixed input energy, fluctuations 
in the PCOPO can
also be $increased$ by the modulation of the refractive index introduced by the PC (Fig.~\ref{fig1}
dot-dashed lines). In other words, when the $signal$ detuning $\Delta_1$ is modulated, the intensity
fluctuations are lowered \cite{Noda2007}, while when (also) the $pump$ detuning $\Delta_0$ is modulated,
such fluctuations increase, in spite of being in the PC band-gap. 

The possibility to either lower or increase the fluctuations of the most intense spatial modes  is due to
the presence of wave-mixing between different frequencies in the parametric oscillator.
Indeed, the fluctuations strength is inherently related to the proximity to the instability threshold  in a
nonlinear system driven out of equilibrium and approaching this point. 
Large fluctuations  are a  clear signature  of an instability and this allows a consistent description of our
results.  In Fig.~\ref{fig1}a, analytical average intensities in the linear approximation~\cite{PCOPObelow} are
shown for the PCOPO below threshold, in different configurations. When the  threshold is raised  (pattern
inhibition due to the PC~\cite{gomilaPRL,damiaPRE}), the fluctuations strength is reduced, while --for the same
 input energy-- if the threshold is lowered by introducing the PC then fluctuations increase.
What we actually find is that, in presence of nonlinear mixing between different fields, the way in which
the PC changes the instability of the signal is not trivial. We remind that when the stripe pattern at the
critical $k_c$ appears in the signal field, the pump also develop a modulation at $2k_c$.  Then for the
detuning modulation $k_{pc}=2 k_c$ the unstable signal wavenumber is in the band-gap but the detuning
profile  somehow `stimulates' the nonlinearly emerging pattern in the pump field. Therefore this PC
modulation provides two competing mechanisms, inhibiting the signal spatial instability, as in
Refs.~\cite{gomilaPRL,damiaPRE}, but also imprinting in the pump the nonlinear structure favoring the
instability process. The vertical asymptotes in Fig.~\ref{fig1}a, where the linear approximation for
analytical calculations breaks down, show  the different threshold values for $E$ depending on the PCOPO
configuration. Notably, if the PC modulates the pump detuning, the parametric threshold can be crossed for
values even $lower$ than in the case of perfectly resonant OPO. 

Apart from the strength of spatial fluctuations, an important aspect
is the quantumness of the correlations. Non-classical effects in multimode OPO are
known to exist between opposite far field modes $+k$ and $-k$ due to emission of
photons pairs in the parametric down-conversion process
\cite{lugiato,opos}. In particular, two-modes
squeezing 
is studied considering the generic joint quadrature $\theta $
\begin{eqnarray} \label{Eq:sigma} 
\Sigma_{\theta\phi}(k,-k)=(\hat A_1 (k) + \lambda \hat A_1 (-k)e^{i\phi} )e^{i\theta} + h.c.\label{quad}
\end{eqnarray} 
with $\phi$ relative phase between the superposed spatial modes. Here we take $\lambda=1$.
\begin{figure}[h]
\includegraphics[width=4cm]{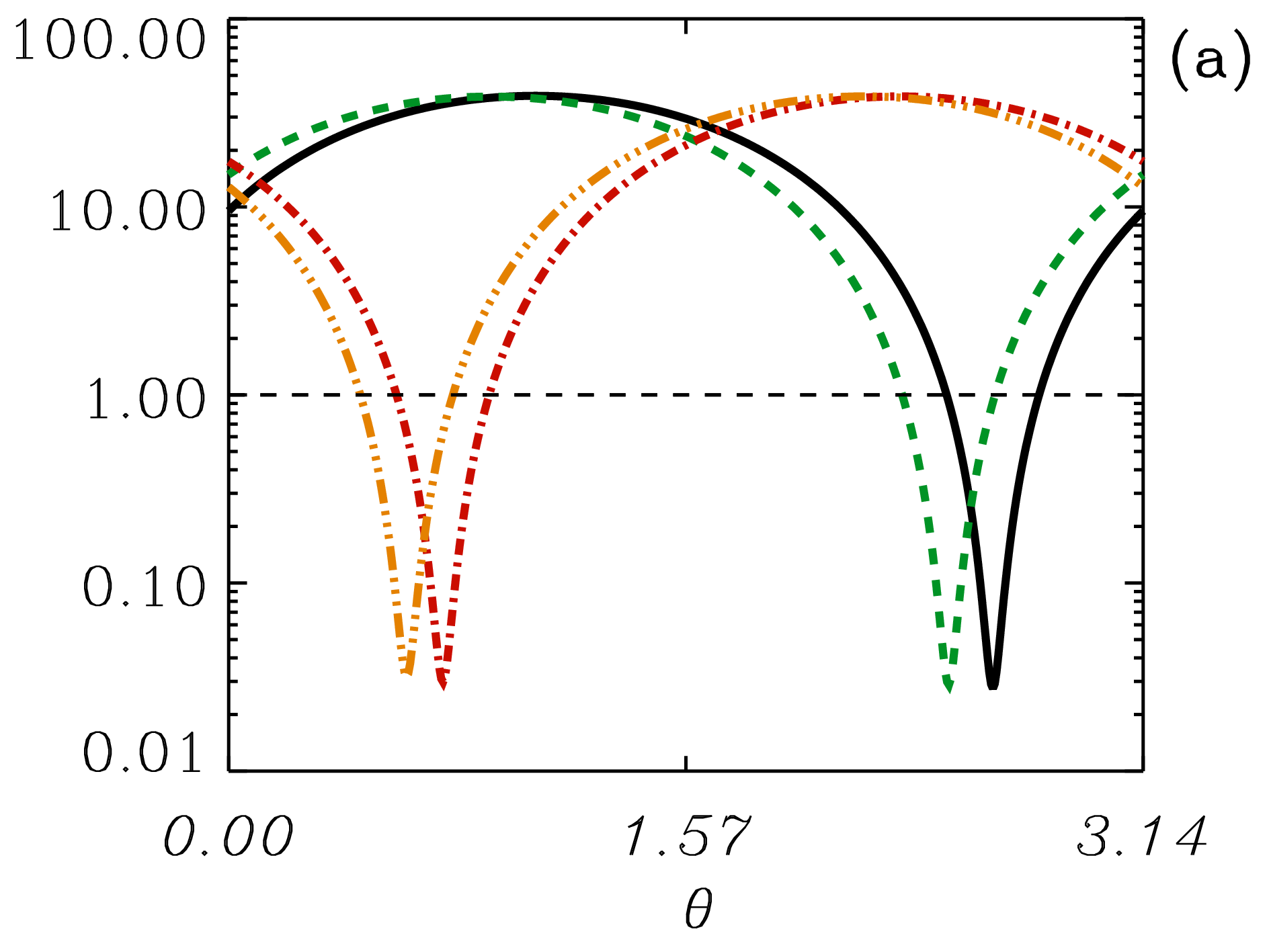} 
\includegraphics[width=4cm]{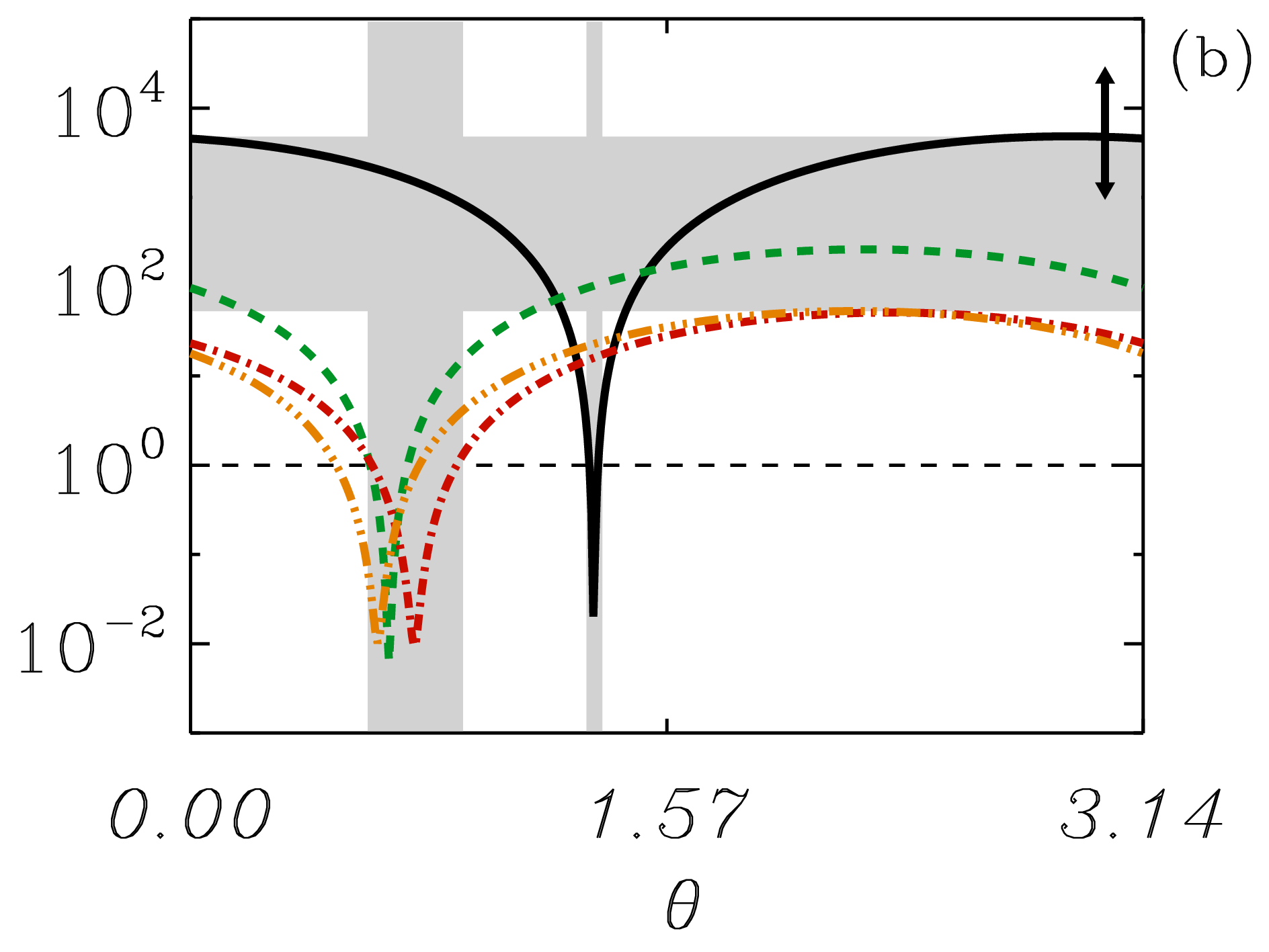} 
\caption{(Color online) Variance of $\Sigma_{\theta\bar\phi}$ (from Eq.~\ref{Eq:sigma} with $\lambda=1$)
as a function of the quadrature angle and for the superposition angle $\bar\phi$ giving
the largest squeezing for each OPO and PCOPO configuration. Pump field $5\%$ below threshold (a) and
$2\%$  above threshold (b). The horizontal dashed line is the shot noise and other 
lines as in Fig.~\ref{fig1}. The small arrow shows the deviation between different 
numerical runs in the OPO, due to phase diffusion. \label{fig2}}
\end{figure}
Squeezing achieved below threshold increases with the pump intensity being maximum at the
parametric threshold \cite{lugiato,opos,zambriniEPJD}. 
Due to the discussed PC effect on  the parametric threshold, the
squeezing attained in OPO and PCOPO will be compared at the same distance from the
respective thresholds. We find then that squeezing achieves similar
values in the OPO and in the PCOPO  modulating pump and/or signal detuning, as shown in
Fig.~\ref{fig2}a, being the major difference in the dependence on the angles 
$\theta$ and $\phi$. 

A different scenario is found $above$ threshold, considering squeezing between {\it intense} modes. In
Fig.~\ref{fig2}b the variance of  $\Sigma_{\theta\phi}$ (Eq.~\ref{Eq:sigma}) for the OPO (upper line) is
compared with the PCOPO (three lower lines). Even if the attained squeezing (minimum value of the plotted
variance) is similar in all cases, there are important differences in the noise present in the unsqueezed
quadrature (maximum value). Far from being in a minimum uncertainty state, the OPO displays an extremely large
noise in the unsqueezed quadrature (black line in  Fig.~\ref{fig2}b) due to the well-known  phase diffusion
between down-converted modes and to excess noise in their relative phase \cite{reid}. In
spatially multimode devices, the visible effect is a  diffusive motion of the excited pattern and has been
related to  translational symmetry break and noise excitation of the corresponding neutral
Goldstone mode \cite{zambrini2000}. On the
other hand,  due to the refractive index modulation, the PCOPO does
not exhibits translational symmetry and the formed pattern is locked to the position of the PC. This leads
to a strong reduction of noise in the PCOPO: in  Fig.~\ref{fig2}b  there are two orders of magnitude between
the variances of the unsqueezed quadrature in the PCOPO  for modulated $\Delta_0$  and the OPO, as
highlighted by the horizontal gray stripe.  An important consequence  in view of applications is that the
reduction of fluctuations in unsqueezed quadratures leads to a significant increase of the range of 
quadratures with sub-shot-noise fluctuations, as highlighted by the two vertical stripes in Fig.~\ref{fig2}b. A
PCOPO will indeed be  more  robust to changes in the choice of the local oscillator phase  $\theta$
as more quadratures are actually squeezed.


\begin{figure}[h]
\includegraphics[width=4cm]{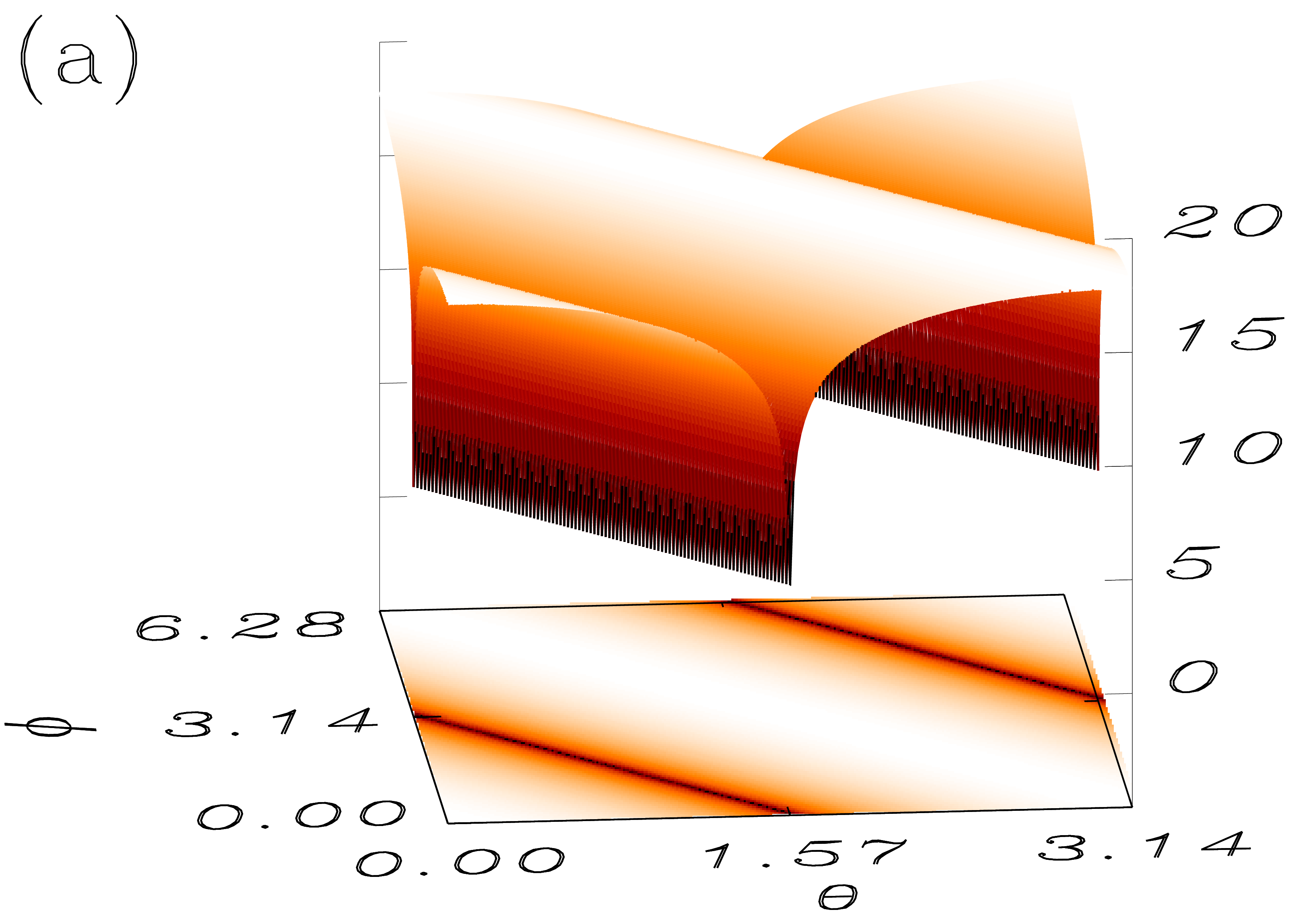} 
\includegraphics[width=4cm]{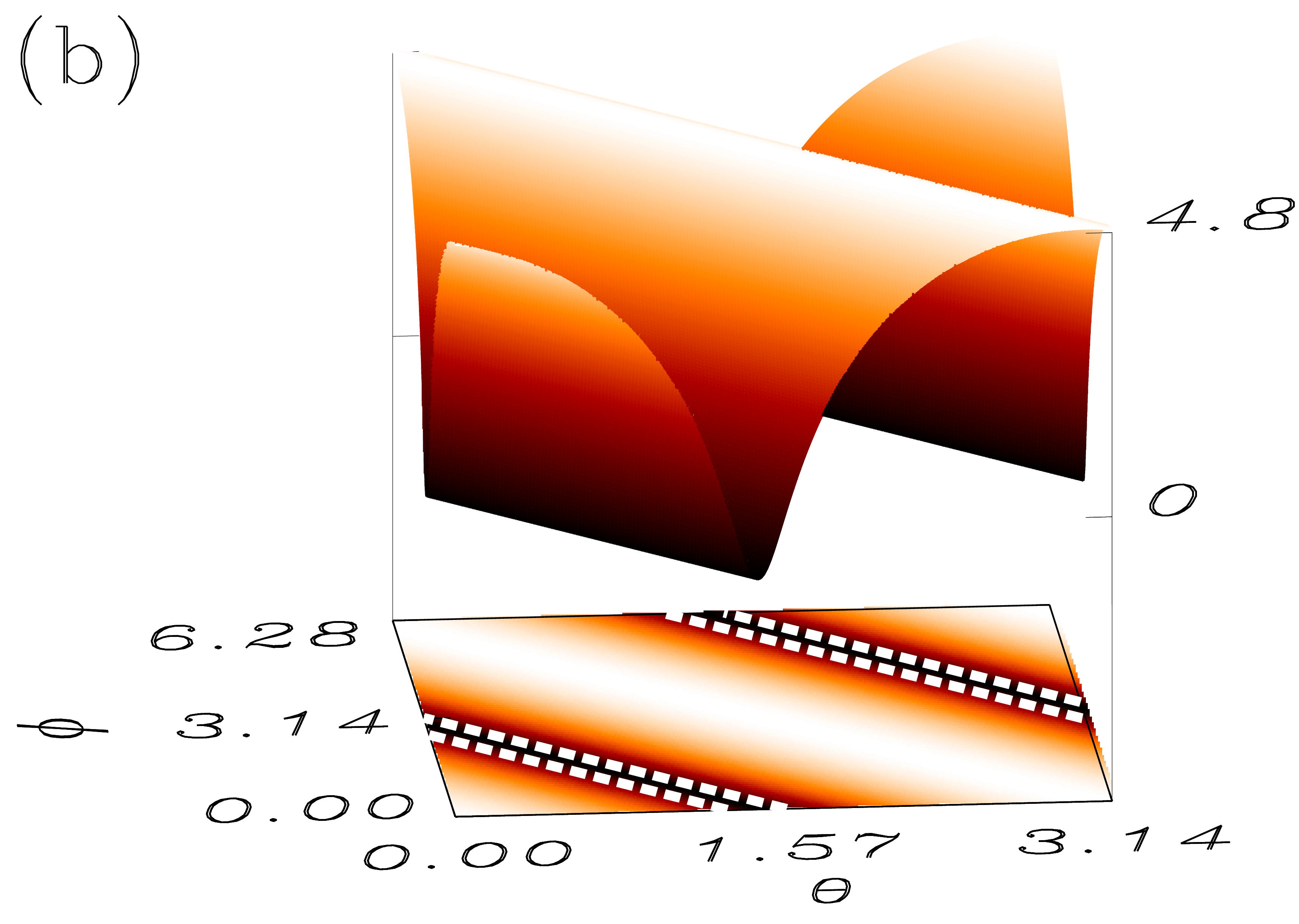} 
\includegraphics[width=4cm]{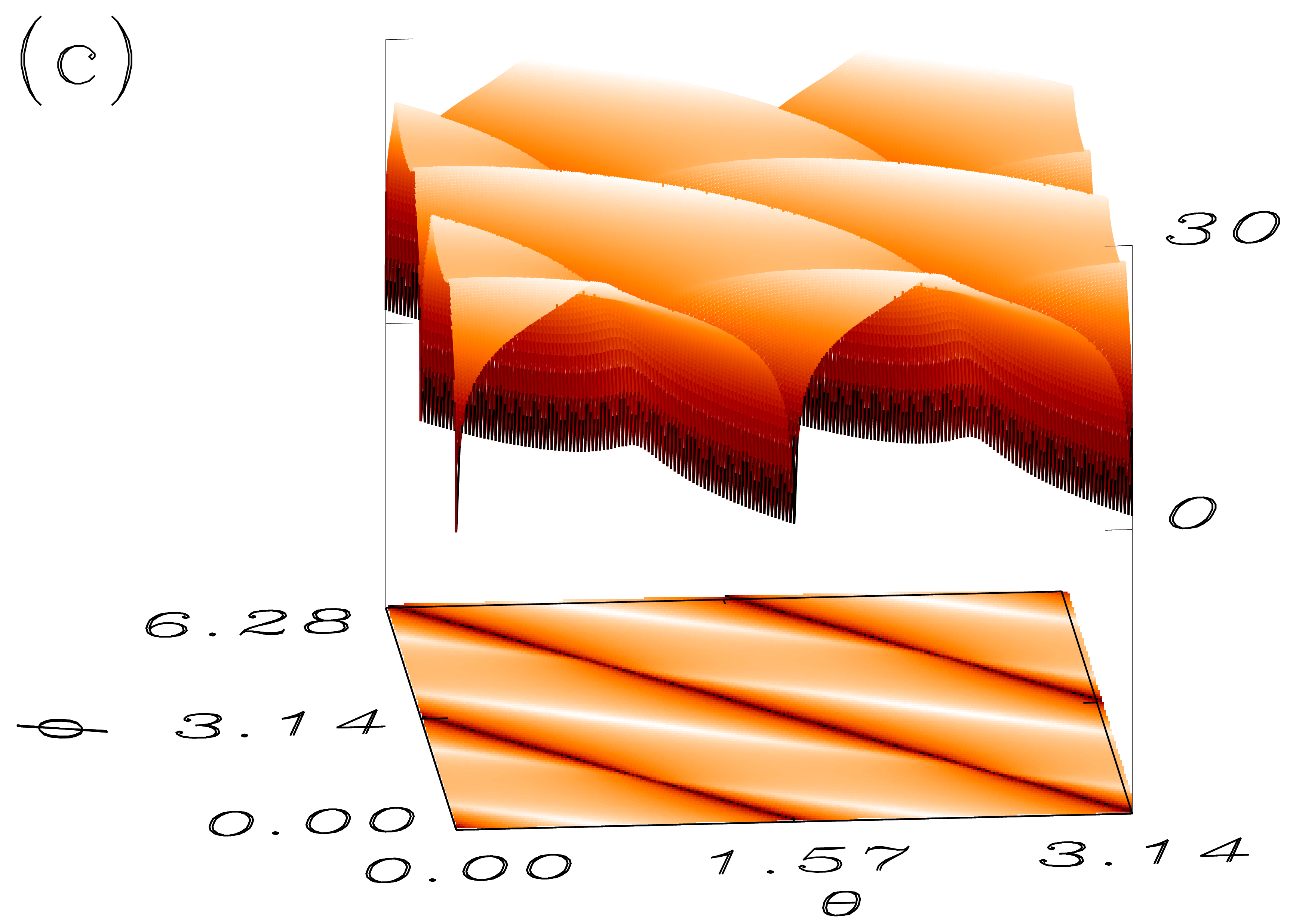} 
\includegraphics[width=4cm]{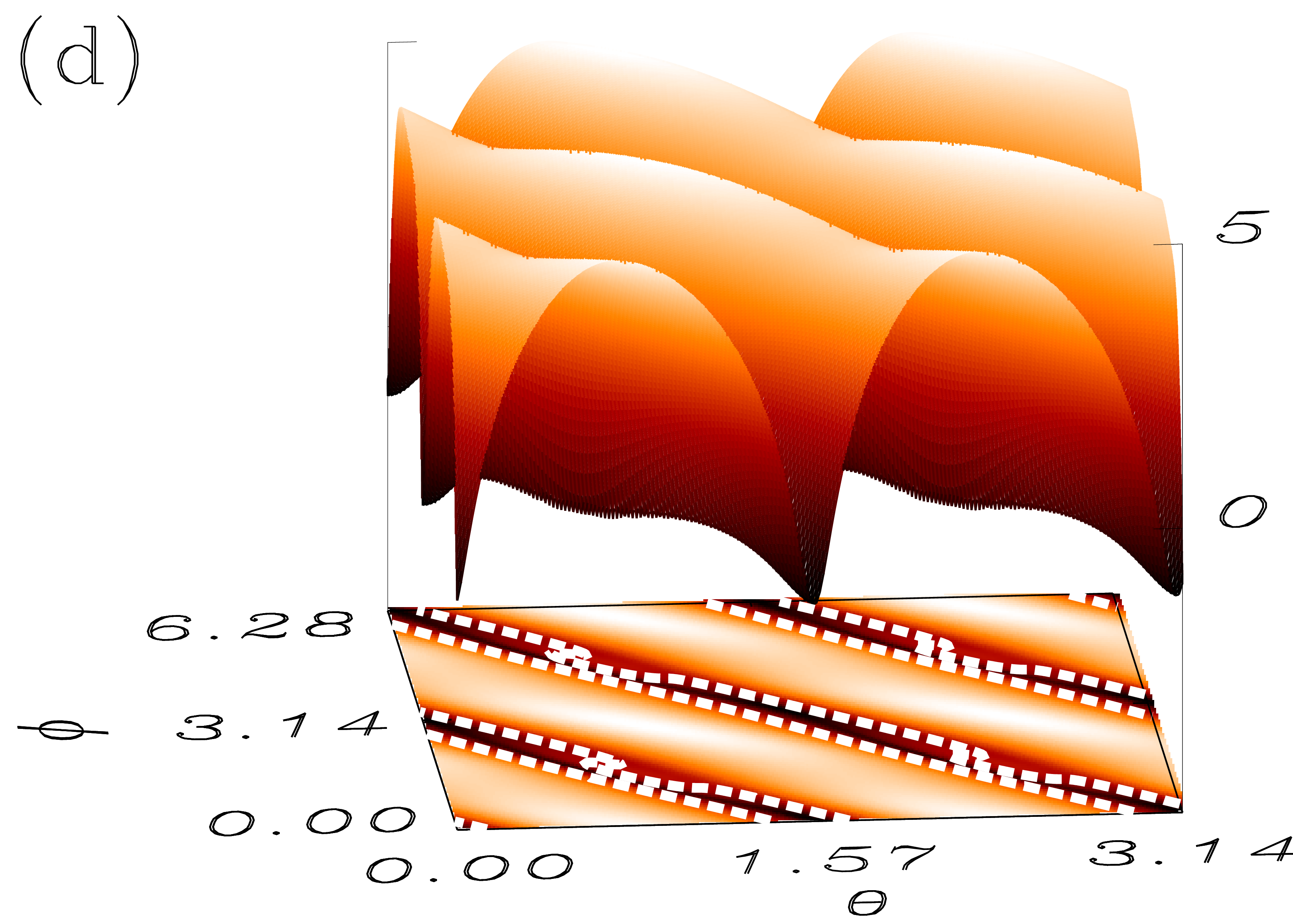} 
\caption{
Variances sum $\mathcal{I}$ (a and b) and product $\mathcal{E}$ (c and d), as defined in Eqs. 
(\ref{duaninequality})  and (\ref{reidinequality}) for the output fields \cite{collett84} of OPO (a
and c) and of PCOPO (b and d) when varying the  superposition and interference angles
$\theta_0,\phi_0$. In (b) and (d) the PC modulates the pump, $\Delta_0=0.5\sin(k_{pc}x)$ (like for
the red dot-dashed line in Fig.~\ref{fig1} and \ref{fig2}) and all results are for pump $2\%$ above
threshold. The white dashed lines limit wide regions regions for which EPR entanglement and
inseparability are respectively predicted.  \label{fig3}} \end{figure}

Multimode OPOs allow to generate not only squeezed but also $spatially$ entangled states 
\cite{opos,eprspatial}. A series of key experiments recently demonstrated  spatial entanglement between
light beams (continuous variable regime) in different optical nonlinear devices \cite{qimaging}.  Here we
show how entanglement in parametric oscillators is changed by the presence of a photonic crystal,
considering two well-known criteria \cite{reid,simon-duan}. One distinguishes states exhibiting
Einstein-Podolsky-Rosen (EPR) paradox  \cite{EPR35} for conditional variances such that
\begin{equation}
\mathcal{E} = \Delta^2\Sigma_{\theta_0,\phi_0}\Delta^2\Sigma_{\theta_0+\frac{\pi}{2},
 \phi_0+\pi}\leq1\label{reidinequality},
\end{equation}
for some  choice of superposition and interference angles $\theta_0,\phi_0$ \cite{reid}. 
Notice that here the
parameter $\lambda$ minimizes each variance of the joint quadrature (\ref{Eq:sigma}), as also
reviewed in Ref. \cite{Reidreview}. A second measure we consider is the inseparability condition
\begin{equation}
\mathcal{I} = \Delta^2\tilde{\Sigma}_{\theta_0,\phi_0}+ \Delta^2 \tilde{\Sigma}_{\theta_0+\frac{\pi}{2},\phi_0+\pi}
\leq 2(a^2+\frac{1}{a^2}).\label{duaninequality}
\end{equation}
with $\tilde{\Sigma}_{\theta,\phi}= (a\hat A_1 (k) + a^{-1}  \hat A_1 (-k)e^{i\phi} )e^{i\theta} + h.c.$ and
positive parameter $a$ \cite{simon-duan}. Below threshold, the presence of the PC changes significantly the
intermode correlations of this systems leading to new not vanishing terms with respect to the case of the
OPO (for instance in the PCOPO $\langle\hat A_1^2 (k)\rangle\neq 0$), but as for squeezing, similar results
are found for the attained entanglement. The most significant differences are found, again, $above$
threshold where the presence of the PC enhances quantum effects leading  to a spatially entangled state. In
this regime, the mentioned phase diffusion \cite{reid,zambrini2000} leads to spikes at low frequency in
noise spectra, preventing entanglement in the OPO, as we show in Fig.~\ref{fig3}a and c. On the other hand,
for a non translational invariant system such as the PCOPO,  we find significant regions in which both EPR paradox
(\ref{reidinequality}) and state inseparability (\ref{duaninequality}) are predicted. 
We stress that in Fig.~\ref{fig3}, OPO and PCOPO are compared at the same
distance form the instability threshold and that, even if the best performance is obtained when 
pump detuning is modulated (Fig.~\ref{fig3}b and d) we find that all
configurations of the PCOPO show some entanglement, degraded when removing the PC. 

 Summarizing, we have analyzed the effect of a PC in a multimode OPO whose detuning suffers a transverse
spatial modulation, in the case in which the spatially unstable mode appearing at threshold falls
within the band-gap.  Due to the presence of nonlinear mixing between waves at $2\omega$ and $\omega$ there
are competing phenomena leading to either inhibition and enhancement of intensity quantum fluctuations for a
fixed pump below
threshold. The possibility to control with the PC the  intensity of quantum fluctuations is related to the
raise (pattern inhibition) and lowering of the parametric threshold.  We find that,
below threshold, the attained spatial squeezing as well as entanglement in the signal field
are preserved  at a fixed distance form the parametric threshold. Notably, above threshold, 
the break of the translational
invariance due to the PC provides a strong mechanism to reduce (up to two orders of magnitude) the quantum
fluctuations associated to pattern diffusion leading to squeezing  
over a significantly larger range of quadrature angles. This would reduce the sensitivity of the
measurable squeezing with the choice of the phase of the local oscillator. 
Moreover, the strong spatial locking due to the presence
of the PC in the OPO allows to generate inseparable as well as EPR entangled spatial beams.
This analysis, restricted to one transverse dimension, is expected to give similar results
when extended to 2D for PC modulated only in one direction, as also confirmed by our analytical results 
below threshold \cite{PCOPObelow}. More complex is the case of a PC with
transverse hexagonal or square geometries where the same pattern selection
process is  an open question.

\acknowledgments{Funding from  FISICOS (FIS2007-60327) and CoQuSys (200450E566) projects is acknowledged. M.A.G.M
acknowledges support by the Fulbright Commission and FECYT.}

\end{document}